\documentclass[aps,prd,superscriptaddress,letterpaper,preprintnumbers,asmmath,amssymb]{revtex4}
\usepackage{amssymb,amsmath,amsbsy}
\usepackage{mathrsfs}
\usepackage[dvips]{graphics}
\usepackage{epsfig}

\def\be{\begin{eqnarray}}
\def\ee{\end{eqnarray}}
\def\bea{\begin{eqnarray*}}
\def\eea{\end{eqnarray*}}

\def\tilde{\widetilde}

\def\centeron#1#2{{\setbox0=\hbox{#1}\setbox1=\hbox{#2}\ifdim
\wd1>\wd0\kern.5\wd1\kern-.5\wd0\fi
\copy0\kern-.5\wd0\kern-.5\wd1\copy1\ifdim\wd0>\wd1
\kern.5\wd0\kern-.5\wd1\fi}}
\def\ltap{\;\centeron{\raise.35ex\hbox{$<$}}{\lower.65ex\hbox{$\sim$}}\;}
\def\gtap{\;\centeron{\raise.35ex\hbox{$>$}}{\lower.65ex\hbox{$\sim$}}\;}

\newcommand{\newc}{\newcommand}
\newc{\qbar}{{\overline q}}
\newc{\Kahler}{K\"ahler }
\newc{\deltaGS}{\delta_{\rm GS}}

\begin{document}
\preprint{
\vbox{\vspace*{2cm}
      \hbox{UCI-TR-2012-24}
      \hbox{May, 2012}
}}
\vspace*{3cm}

\title{Pseudo-Higgs Signals at the LHC}
\author{Linda M. Carpenter and Jessica Goodman}

\affiliation{Department of Physics and Astronomy  \\
   University of California Irvine, Irvine, CA U.S.A. \\
   lcarpent@uci.edu, j.goodman@uci.edi  \\
\vspace{1cm}}

\begin{abstract}

We consider general fermi-phobic scenarios in which excess events in
diphoton or WW/ZZ resonances may be seen at LHC.  These Higgs like signals
do not necessarily suggest that the new resonance is a particle with Yukawa
couplings nor do we know that it is responsible for electroweak symmetry
breaking. We can, however, extract two facts from it, this particle couples
to pairs of SU(2) and U(1) gauge bosons and it must be a scalar,
pseudoscalar, or tensor.  We consider the signals of general operators up to
effective dimension 5 in which a new scalar, psuedo-scalar, or tensor
particle may couple to pairs of standard model gauge bosons.  This particle
may or may not be charged under the standard model gauge groups, and may be
produced via gluon fusion or EW vector boson fusion.

\end{abstract}

\pacs{}
\maketitle

\section{Introduction}
The Standard Model (SM) has been amazingly successful in describing fundamental interactions however, it remains
incomplete.  The Higgs boson is most likely the particle responsible for electroweak symmetry breaking, though it has yet to be discovered.  The Higgs mass is not a known quantity however, its production cross sections and branching fractions to SM particles are well known \cite{Dittmaier:2011ti}\cite{Dittmaier:2012vm}\cite{Dawson:1998yi}.  At the LHC both the production and the decay of the Higgs boson are dominated by processes in which the Higgs has effective couplings to pairs of SM gauge bosons.   While Higgs production is dominated by gluon fusion \cite{Baglio:2010ae}, it does not couple directly to gluons but has a large effective coupling them through a loop of heavy quarks.   At low masses $(m_H \lesssim 130 \text{GeV})$, the Higgs decays predominately  to $b\bar{b}$ however, large QCD backgrounds at LHC make this channel difficult to see.  Thus, the Higgs decays to pairs of electroweak gauge bosons are crucial for Higgs searches.  In fact one of the most important channels for Higgs discovery is  $h\rightarrow \gamma \gamma$.  Again, the Higgs does not couple directly to photons but it has an effective coupling brought about by loops of charged W bosons and heavy third generation quarks. In fact, decays to weak boson pairs become dominate once we reach the kinematic threshold since the leading term in the partial width of Higgs to diboson is cubic in Higgs mass versus linear in Higgs mass for partial width to fermion pairs.

LEP, Tevatron, and now ATLAS and CMS have been chipping away at Higgs mass parameter space leaving only
a small window for the SM Higgs to hide \cite{TEVNPH:2012ab}\cite{Barate:2003sz}.  Recent search results from both ATLAS and CMS have reported a small excess in the WW channel and a slightly larger excess in the $ZZ$ and $\gamma\gamma$ channels around $125\text{GeV}$ \cite{:2012si}\cite{Chatrchyan:2012tx}.  While this excess may be the first hint at the SM Higgs, it is  not yet a discovery.

However, seeing a mass resonance in a diboson channel does not necessarily mean that the particle being produced is a Higgs.  One would still be lacking evidence that the particle has the standard Yukawa couplings, or that it is responsible for electro-weak symmetry breaking.  In fact, if a mass resonance is seen in and EW boson channel the new state may not have effective couplings to gluons as the Higgs does, unlike the Higgs whose production proceeds dominantly through gluon fusion, the new state may be dominantly produced through electro-weak couplings.  A Higgs-like diboson signal may be the consequence of a variety of effective operators where a new state couples to pairs of standard model bosons.  Such a particle may a scalar, pseudoscalar, or tensor; it may or may not carry standard model quantum numbers; and it need only have large enough effective couplings to $SU(2) \times U(1)$ gauge bosons and may or may not couple to gluons. Various scenarios have been studied exploring some of these possibilities \cite{Gabrielli:2012yz}\cite{Ellis:2012wg}\cite{Low:2011gn}.

In this paper we present various Pseudo-Higgs scenarios in which a particle with effective couplings to $SU(2) \times U(1)$ gauge bosons may mimic a Higgs.  We attempt to study general operators for all fermi-phobic scenarios up to effective dimension 5.   We explore constraints on the available parameter space and explore which regions of this space might be responsible for the observed di-photon excess at LHC.  We pay particular attention to the  possibility that new states may be produced dominantly by weak-boson fusion.  The paper proceeds as follows, in Section 2  we consider the lowest dimension effective operators in which a scalar field may couple to pairs of  $SU(2) \times U(1)$ gauge bosons- both the 'bosonic Higgs' scenario and a mimic scenario in which a scalar field couples pairs of gauge bosons through a modified Higgs kinetic term.  In Section 4 we consider couplings of a SM singlet to gauge bosons through effective dimension 5 operators.  We consider both pure scalar and pure pseudoscalar fields which are singlets under all standard model gauge groups.  In section 5 we consider effective operators coupling a spin 2 field to pairs of SM gauge bosons.

\begin{figure}[h]
\includegraphics[width=7.8 cm]{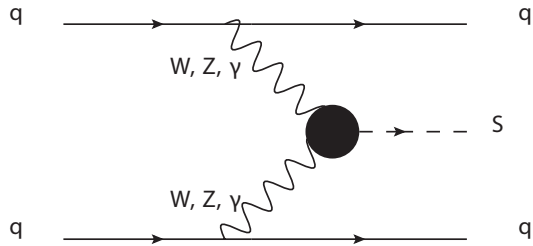}
\caption{Production of a SM singlet state through weak boson fusion}
\end{figure}

\section{Review of Bosonic Higgs}
Here we review the `Bosonic Higgs', a non-standard Higgs scenario in which a particle is produced and decays to pairs of electroweak bosons thus mimicking a standard Higgs signal.  In the model under consideration a field exists which has the quantum numbers of the standard model Higgs. The field is an SU(2) doublet which gets a vacuum expectation value breaking the $SU(2)\times U(1)$ gauge group.  However, this field does not have any Yukawa couplings.  At tree level the Higgs-like particle couples only to W and Z bosons and the coupling occurs through the field's kinetic term.

\bea
DH(DH)^{\dagger}=((\partial- gA)H)^2
\eea

We see that by expanding the covariant derivative one may take the piece proportional to $A^2$ and insert one power of the Higgs vev getting a term in the Lagrangian
\bea
L \sim vHAA^{\dagger}
\eea

This is a term of effective mass dimension 3 coupling the Higgs-like field to the W and Z gauge bosons.  Since there are no Yukawa couplings to couple it to heavy quarks this field has no effective couplings to gluons.  The field does have an effective coupling to photons which it gets through loops of charged W bosons.

Collider consequences of such a scenario have been considered by \cite{Akeroyd:1995hg}\cite{Abbott:1998vv} and recently by \cite{Gabrielli:2012yz}.

In this case Higgs production proceeds dominantly through weak-vector boson fusion.  For a Higgs with masses under 160 GeV, decay is dominated by the process $h\rightarrow WW^{*}/ZZ^{*}$.  However the Higgs may decay to pairs of photons through a loop of charged W bosons.  The Higgs decays to $Z\gamma$ through a similar process.  The total branching fraction of the Higgs to photons in this scenario is much larger than in a standard Higgs scenario.  One reason for this is due to the lack of Yukawa couplings, this ensures that the branching fraction is not eaten up by decays to lepton pairs and to gluons.  Another reason is because in the standard scenario, the Higgs couples to photons through loops of charged W's but also heavy quarks, which partially cancel the W loops.  Without Yukawa couplings these cancelations do not take place and the partial width to photons is increased (see for example \cite{Rainwater:2007cp}).  Thus, though the total Higgs production cross section in such a scenario is smaller than through a gluon fusion process, the total production cross section of di-photons in this scenario may be appreciable, effectively mimicking a standard Higgs signal in the di-photon channel.

\subsection{New Effective Operators}


\begin{figure}[h]
\includegraphics[width=7.8 cm]{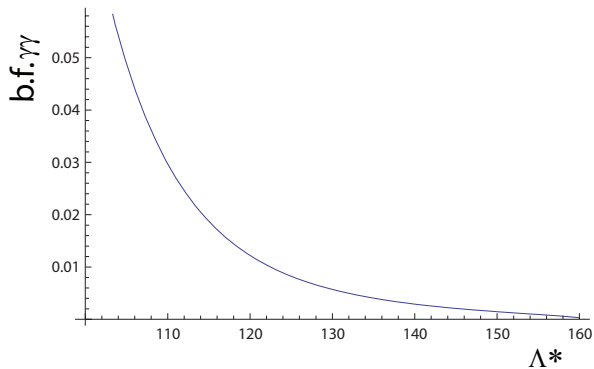}
\caption{Plot of scalar mass vs. branching fraction to di-photons in the Bosonic Higgs scenario}
\end{figure}

We now consider a scenario which has similar physics to that of the Bosonic Higgs case but with a much different particle.   One may consider a particle S, which is a singlet under all gauge groups. One may write an effective coupling of this particle to the Higgs kinetic term which is not forbidden by any symmetry.  The operator is dimension 5 and suppressed by a scale $\Lambda$
\bea
\frac{k S}{\Lambda}DH(DH)^{\dagger}
\eea
Here k is an arbitrary coupling. Now expanding the covariant derivative, extracting the $A^2$ piece we get
\bea
\frac{k^{'} S}{\Lambda}DH(DH)^{\dagger}=\frac{k^{'} S}{\Lambda}((\partial- gA)H)^2\rightarrow \frac{k S}{\Lambda} AA^{\dagger}HH^{\dagger}
\eea

By inserting both Higgs vevs we get a coupling of the scalar field S to pairs of electro-weak gauge bosons.  This effective operator is
\bea
\frac{k v^2}{\Lambda}SAA
 \eea
has the same mass dimensions as the Bosonic Higgs operator but is multiplied by a factor $kv/\Lambda$.  We see that a dimension 5 operator has become an operator of effective mass dimension 3.  Also, in the Bosonic Higgs scenario, the coupling of the Higgs field to pairs of W and Z bosons was completely determined by the Higgs vev and the gauge coupling.  In this scenario, however,  the coupling to the gauge bosons may be varied by shifting the effective cutoff of the theory $k/\Lambda$.  As in the Bosonic Higgs scenario, a coupling to pairs of photons and to $Z\gamma$ will be generated at one loop through couplings to charged W bosons.  When the scalar state decays, its branching fraction will be dominated by decays to pairs of W and Z bosons however, decays to di-photons will make up a few percent of its total branching fraction.  Figure 2 shows a plot of scalar branching fraction into photons vs. the scalar mass for a particular choice of effective cutoff.  The branching fraction of this state to photons is nearly 100 times that of a standard Higgs of the same mass.

\begin{figure}[h]
\begin{center}$
\begin{array}{cc}
\includegraphics[width=7.8 cm]{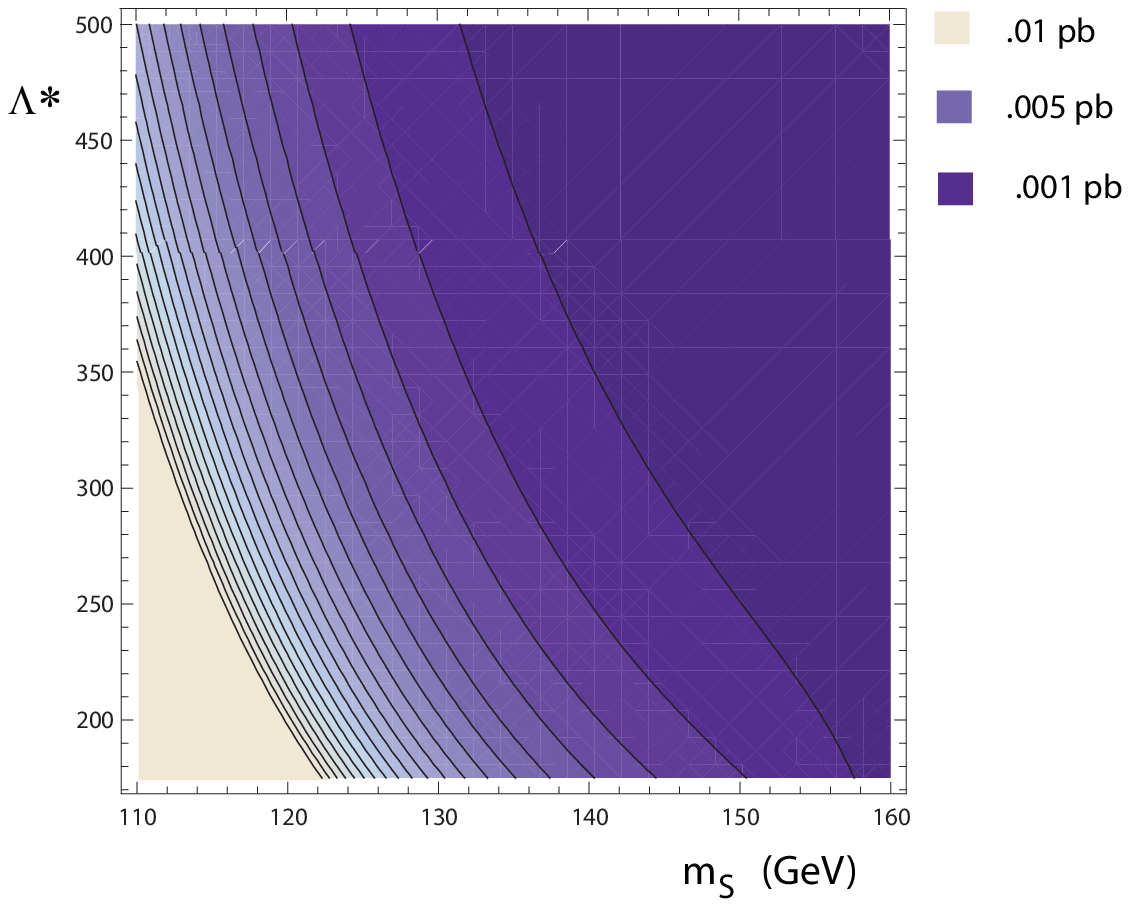} &
\includegraphics[width=7.8 cm]{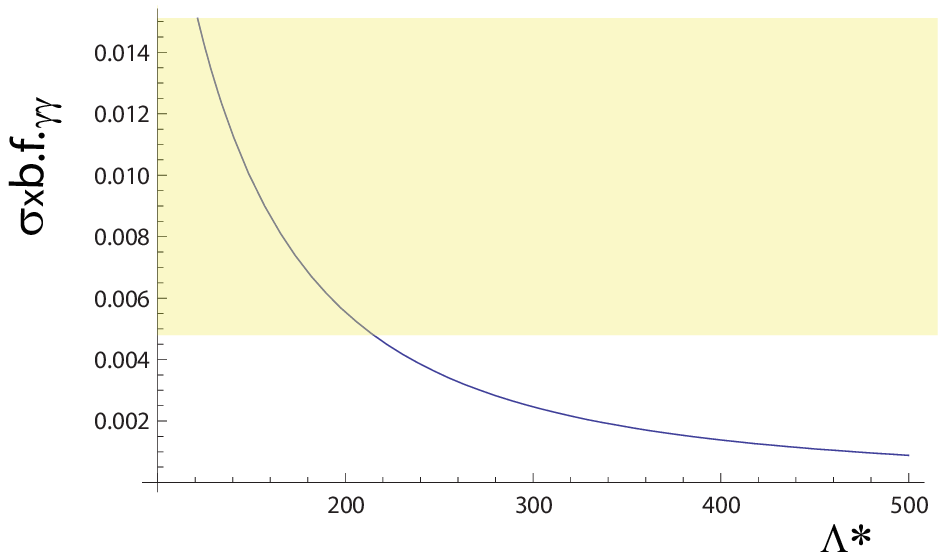}
\end{array}$
\end{center}
\caption{Left- Contour Plot of total di-photon production cross section in the scalar vs. effective cut-off mass plane. Right- allowed region in effective cuttoff parameter space consistant with the LHC di-photon excess for a 125 GeV mass resonance}
\end{figure}

Like the Bosonic Higgs such a scalar state has no effective coupling to gluons.  At LHC, such a state will be produced through the process involving electro-weak vector bosons, one such process is weak-boson fusion as seen in figure 1.  Though this state would only have electro-weak production cross section, its relatively large branching fraction into photons means that it would have an observable signal.  The entire process of production and decay would be  $q q \rightarrow S q q \rightarrow \gamma \gamma q q$.  We may check to see if such a theory is consistent with LHC's observed di-photon excess at 125 GeV.  We simulated the production of di-photon events at LHC for 7 TeV with the process $q q \rightarrow S q q \rightarrow \gamma \gamma q q$.  Events were simulated using MADGRAPH 5 \cite{Alwall:2011uj} and showered with PYTHIA \cite{Sjostrand:2006za}.  It was found that the acceptance for these events was similar to the acceptance of Higgs events decaying to di-photons and produced through the process $g g \rightarrow h$.

Figure 3 a shows the overall production times branching fraction for di-photon events in our scenario vs. the variable effective cut-off scale for a scalar of mass 125 GeV.  One sees that for sufficiently high cut-off the effective coupling of the scalar to SM gauge bosons is suppressed and the total production cross section falls off.  The overall strength of the signal will depend on the scalar mass and on the effective cut-off of the theory.  In figure 3b we present the region in the scalar mass plane vs effective cut-off plane and find it is consistent with the di-photon excess at LHC at 125 GeV.


\section{Dimension 5 operators}
Alternately, one may consider operators of effective dimension 5 through which a particle that is a singlet under all standard model gauge groups may couple to many or all of the SM gauge boson.  In this case the particle may be either a scalar or pseudoscalar.  The Lorentz and gauge invariant operators one may write are
\bea
L=k_i S F^{\mu \nu} F_{\mu \nu}/\Lambda \\ L=   k_i S_p F^{\mu \nu}\tilde{ F_{\mu \nu}}/\Lambda
\label{eq:SMSinglet}
\eea

Where $S$ is a scalar field and $S_p$ is a pseudoscalar. One sees that the scalar couples to the square of the field strength tensor, while the pseudo-scalar, which is parity odd, couples to the dual tensor.  Unlike the Bosonic-Higgs like scenarios here the scalar or pseudo-scalar field may have independent couplings to each SM gauge group and thus there is an independent parameter for each gauge group.

One may consider the parameterization
\bea
k_i^{*} = k_i/{\Lambda}
\eea
where we have absorbed the couplings into the inverse of the cutoff scales. As $k^{*}$ falls the effective scale of the cutoff increases.  These three parameters, $k^{*}_i$ along with the mass of the new particle determine all of the new observable physics of the scenario.  The coupling coefficients of the scalar or pseudoscalar particle to pairs of SM gauge bosons are given by:
\bea
g_{SWW}&=&\frac{2k_2}{sw^2 \Lambda} \\
g_{SZZ} &=& \frac{g_2^2}{4e^2}(\frac{k_1 sw^2}{cw^2}+\frac{k_2 cw^2}{sw^2})/{\Lambda} \\
g_{S\gamma\gamma}&=&\frac{g_1^2}{4e^2}\frac{k_1+k_2}{\Lambda} \\
g_{SZ\gamma} &=& \frac{g_1g_2}{2e^2}(\frac{k_2}{sw^2}-\frac{k_1}{cw^2})/{\Lambda} \\
g_{Sgg}&=&\frac{k_3}{\Lambda}
\label{eq:prefactors}
\eea

Many of these couplings may be varied independently. First we see that the SU(3) coupling is independent from all others.  The new particles may or may not couple to gluons.  The coupling to W's depends only on the SU(2) parameter $k_2$. Thus, the coupling to W's may be turned off or varied while the coupling to photons remain the same by compensating with a change in $k_1$. The couplings to photons and the Z boson may not be independently varied.

\begin{figure}[h]
\begin{center}$
\begin{array}{cc}
\includegraphics[width=5.8 cm]{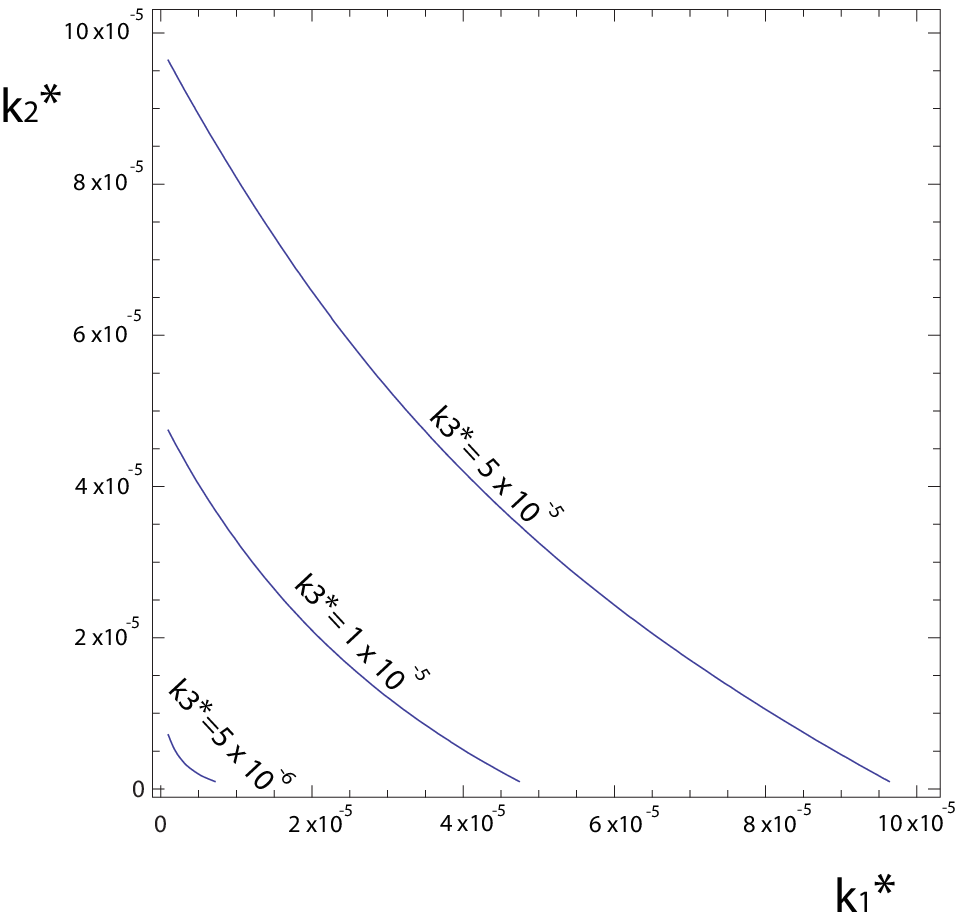}
\includegraphics[width=5.8 cm]{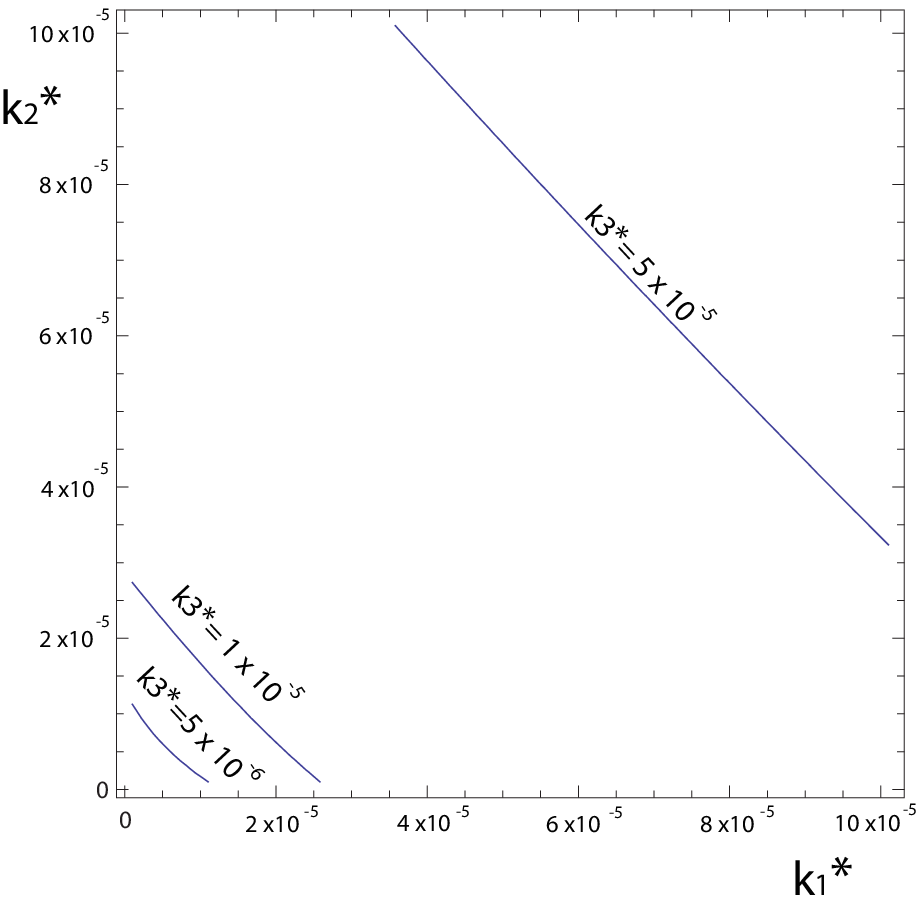}
\end{array}$
\end{center}
\caption{For Left- pseudo-scalar and Right- scalar models: Contour for which the di-photon branching fraction is .1 over effective $SU(2) \times U(1)$ coupling space for varying values of effective $SU(3)$ coupling.  }
\end{figure}

\begin{figure}[h]
\begin{center}$
\begin{array}{cc}
\includegraphics[width=5.8 cm]{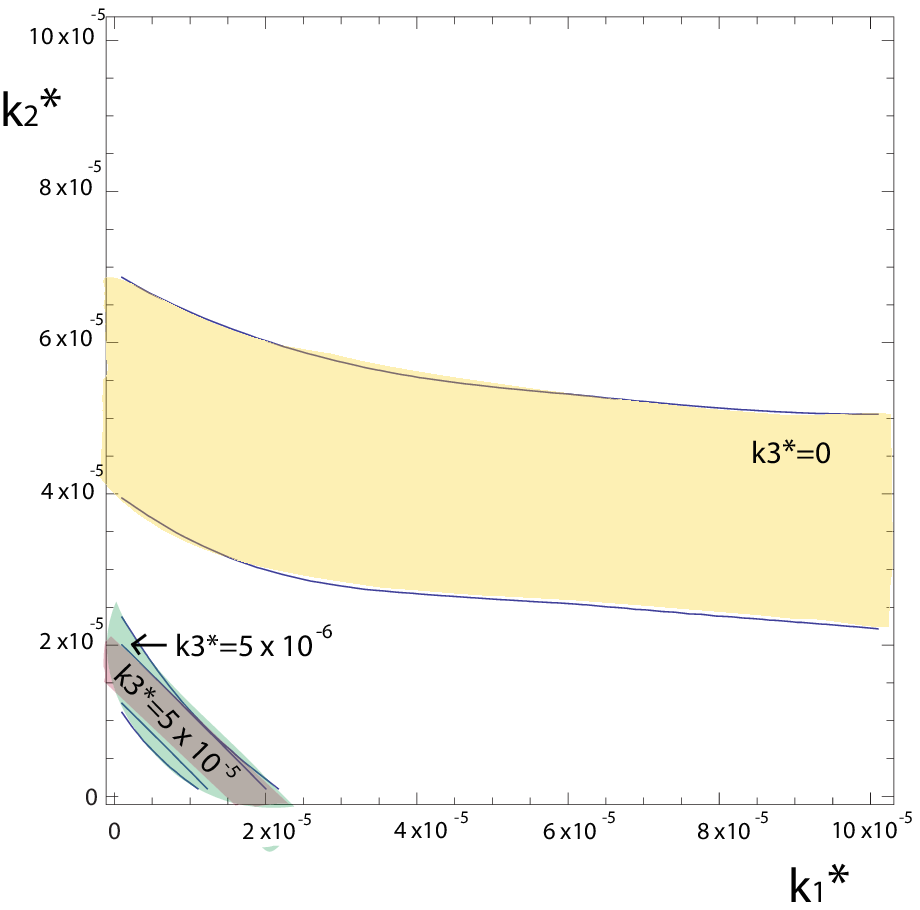} &
\includegraphics[width=5.8 cm]{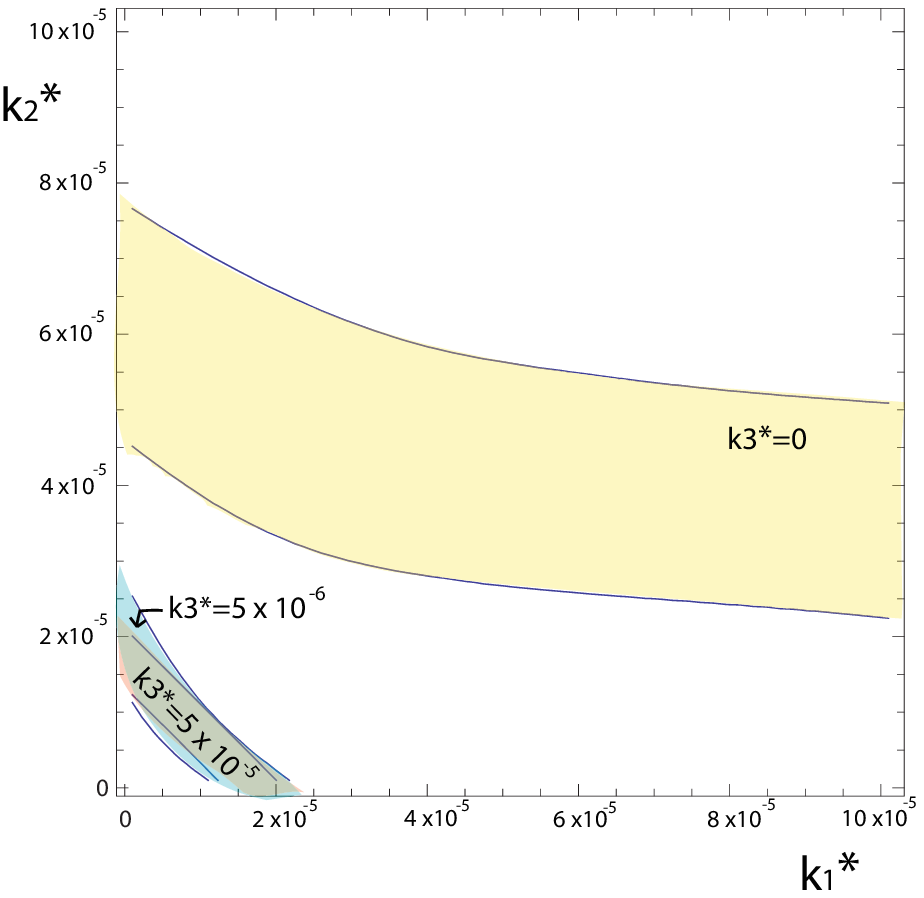}
\end{array}$
\end{center}
\caption{For Left- pseudo-scalar and Right- scalar models with dimension 5 couplings to gauge bosons: regions in effective $SU(2) \times U(1)$ coupling space which fit the LHC di-photon excess for a 125 GeV resonance for varying values of effective $SU(3)$ coupling. }
\end{figure}

Different values of the parameters allow for different modes of scalar or pseudo-scalar  production to be dominant.  If the couplings to gluons are at all appreciable then gluon fusion becomes the most important production mechanism for the new particle-it is produced through the process $g g\rightarrow S$.  However, if the coupling to gluons is not present or very tiny then the production process is dominated by the weak boson processes, $q q \rightarrow S q q$.  One finds in general that the production cross sections of scalar and pseudo-scalar particles are quite similar.

In addition the branching fractions depend heavily on the choice of parameters.  In general, the branching fractions into massive final states, $WW^{*}$ and $ZZ*$ are quite small compared to that of massless final states $\gamma\gamma$ and $g g$.  The branching fraction into di-photon may be large over a wide range of scalar or pseudo-scalar masses if the effective couplings $k_1^{*}$, $k_2^{*}$ are big enough.  If the coupling to SU(3) is zero or negligible then photons dominate the branching fractions.  As the coupling to gluons is turned on then the branching fraction to gluons quickly becomes large.  We have given analytic formulae for both scalar and pseudo-scalar branching fractions in the appendix.

To illustrate the behavior of branching fractions, figures 4 a and b show contours where the photon branching fraction is 10 percent in the space of effective couplings $k_1^{*}$, $k_2^{*}$.  The contours are shown for increasing value of effective coupling $k_3^{*}$.   One sees that at small values of $k_3^{*}$ the branching fraction to photon pairs will remain above 10 percent even for small values of effective coupling $k_1^{*}$, $k_2^{*}$, but as $k_3^{*}$-the effective coupling to gluons gets larger then the effective coupling to the weak bosons $k_1^{*}$, $k_2^{*}$ must get much larger to keep the di-photon branching fraction above ten percent.

Such a model with a scalar or pseudo-scalar may have Higgs-like signals.  Throughout much of parameter space such a scalar or pseudoscalar may produce many di-photon events. Singlet production may proceed either through gluon fusion, or if the couplings to gluons are absent, through weak boson fusion. The entire processes are $q q \rightarrow S q q \rightarrow \gamma \gamma q q$, and $g g \rightarrow S \rightarrow \gamma \gamma$. In figure 5 we present the region of parameter space where a 125 GeV scalar or pseudoscalar may be consistent with LHC's observed excess.  Events were simulated using MADGRAPH 5 \cite{Alwall:2011uj} and showered with PYTHIA \cite{Sjostrand:2006za} while the branching fractions were calculated analytically.  We have shown the allowed regions of parameter space both where the coupling to gluons is zero and where it is finite.  One sees that for appreciable couplings to gluons, the overall production cross section of the singlet field is quite high, therefore in order to be consistent with experiment the di-photon branching fraction must be small, forcing one into a small corner of $k_1^{*}$, $k_2^{*}$ parameter space. Note there is a large region of viable parameter space where the couplings to gluons is zero and all processes are electro-weak.
\begin{figure}[t]
\includegraphics[width=7.8 cm]{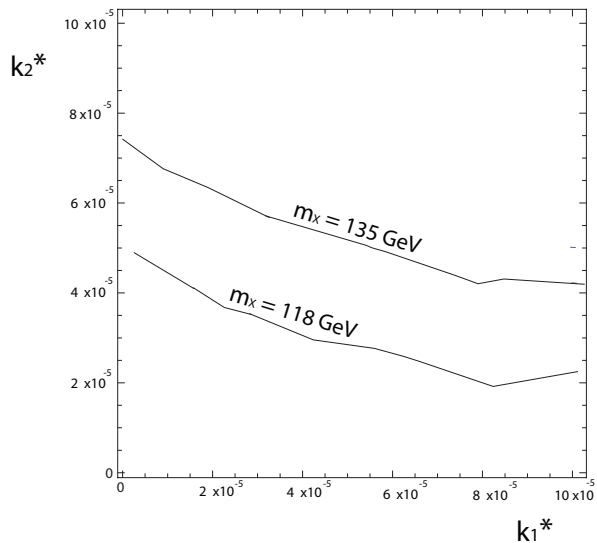}
\caption{For Scalar models with dimension 5 couplings to gauge bosons.  Upper bounds in effective $SU(2) \times U(1)$ coupling space with $SU(3)$ coupling $k_3^{*} = 0$ for various scalar masses. }
\end{figure}

Alternatively, one may look to the current Higgs searches to place bounds on the parameter-space of new scalar and pseudo-scalars, as these states may exist with mass values that have been ruled out for standard Higgs scenarios.  For example one may translate the Higgs exclusions in the di-photon channel to upper limits on the production cross-section of di-photons for new resonances.  Figure 6 for example gives upper bounds in $k_1^{*}$, $k_2^{*}$ parameter space for a scalar particle with dimension 5 coupling to photons for a range of masses.

\section{Operators with Tensors}
Another possibility for a particle which fake Higgs signals by decaying into gauge bosons is a spin 2 state. Such a particle should be a singlet under the Standard Model gauge groups, and of course must not get a vacuum expectation value.  If we are considering operators up to dimension 5 which couple a spin 2 particle to gauge bosons we have  two possibilities

\bea
\mathcal{L} = \frac{\lambda_i}{\Lambda}(T^{\mu\nu} - 4T^{\mu\nu})F_{\mu}^\rho F_{\nu\rho} + \frac{\kappa_i}{\Lambda}(T^{\mu\nu} - 2T^{\mu\nu})m_A^2 A_\mu A_\nu
\eea

The first term is an operator of effective dimension 5 similar to those we  have already considered.  The second term is equivalent to the effective dimension 3 operators we have considered which may be generated by coupling the spin 2 field to a Higgs kinetic term.  These couplings are quite similar to those of a KK gravition to pairs of gauge bosons, see for example \cite{Han:1998sg}.  In the case of the KK graviton both sets of operators must occur together, but we may consider them separately.  In addition, in the case of KK gravitions the coupling to pairs of different standard model gauge bosons must be universal, while in out more general case we are not under such a constraint.

The first operator allows a spin two state to couple to all pairs of gauge bosons, $gg$, $WW/ZZ$, $\gamma \gamma$ and $Z \gamma$.  However, the second operator only allows the spin two state to couple to W and Z bosons, the coupling of the spin two state to photons must then proceed through a loop of W bosons similar to the bosonic Higgs.  Because the effective coupling to photons in that case is of higher dimension, we will here only analyze in depth the phenomenological consequences of the first operator.

\begin{figure}[h]
\begin{center}$
\begin{array}{cc}
\includegraphics[width=9.5 cm]{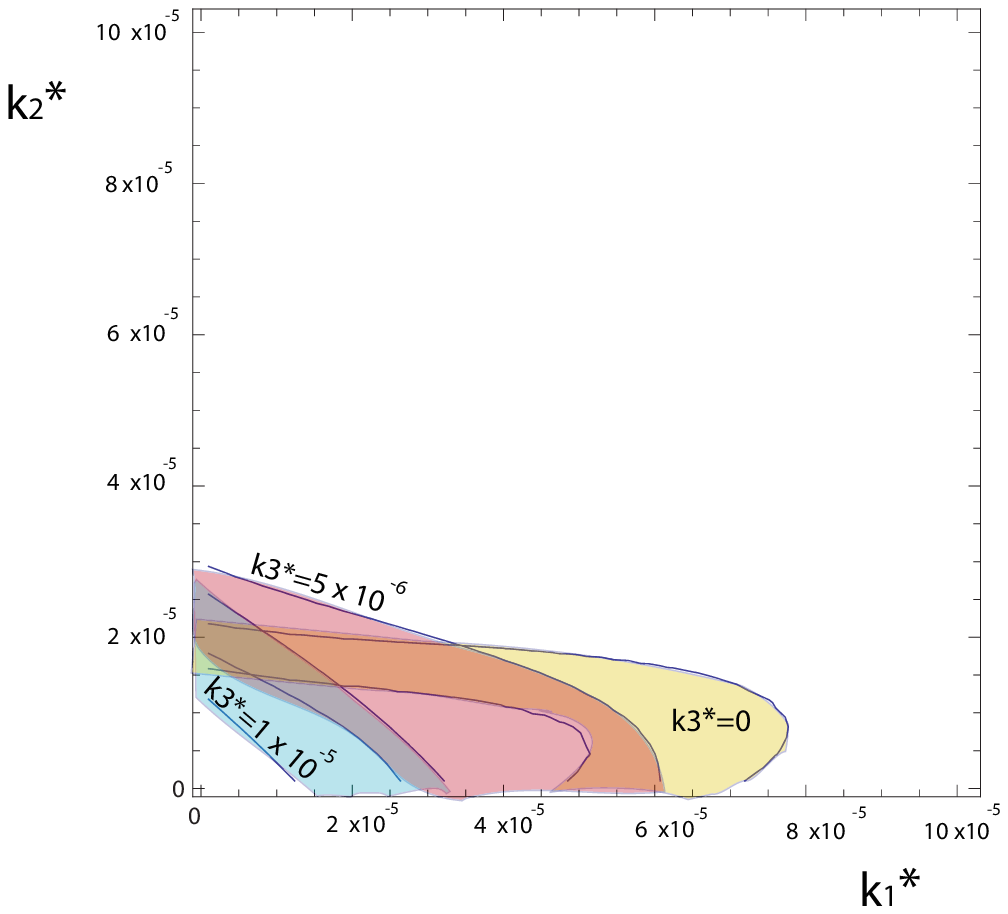} &
\includegraphics[width=8.5 cm]{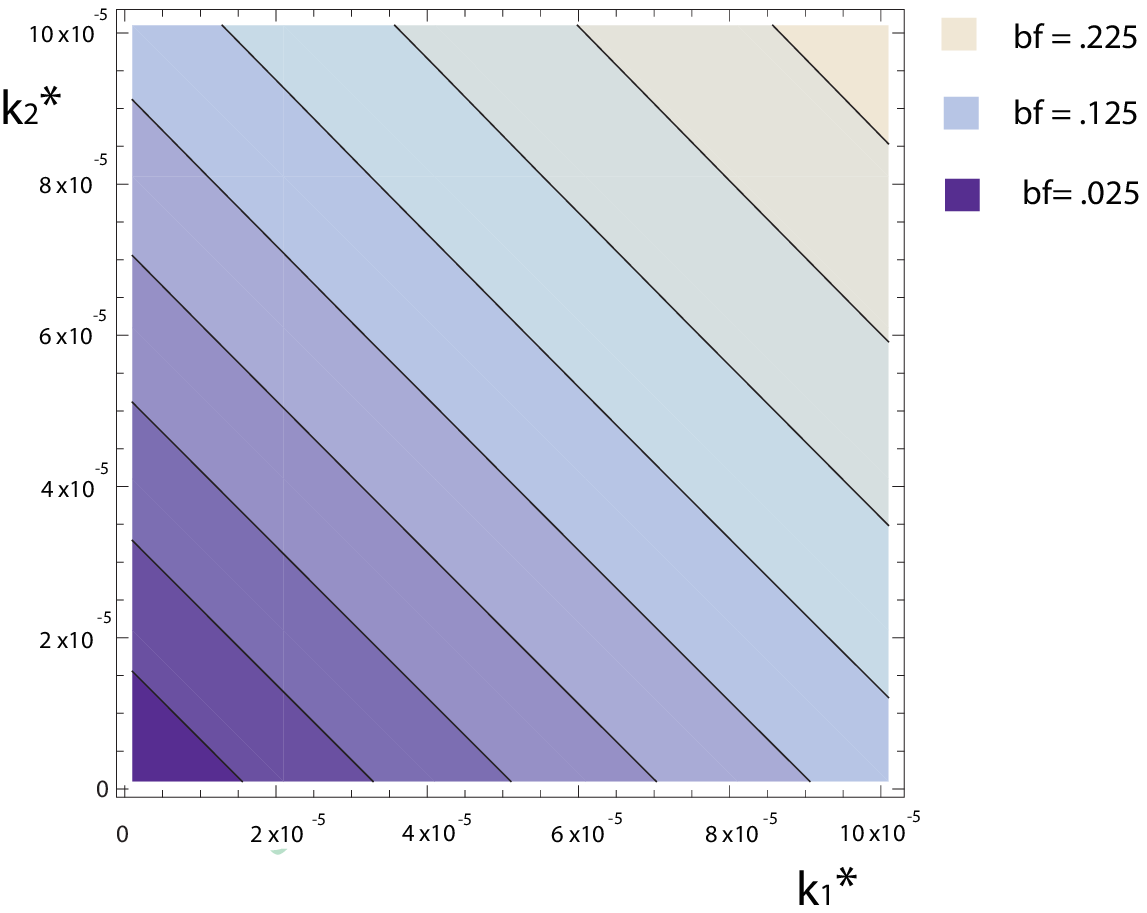}
\end{array}$
\end{center}
\caption{For models of a tensor pseudo-higgs with coupling to SM gauge bosons: Left) Regions in effective $SU(2) \times U(1)$ coupling space which fit the LHC di-photon excess for a 125 GeV resonance with varying values of effective $SU(3)$ coupling. Right) Contour plot of branching fraction into di-photons in effective $SU(2) \times U(1)$ coupling space for effective $SU(3)$ coupling $k_3^{*} = 10^{-5}$ }
\end{figure}

We see that the coefficients of the couplings to pairs of gauge bosons is the same as those given in section 3.  The couplings to gluons and W pairs may be varied separately.  Once again if there is no couplings to gluons then the production of the spin two state proceeds through weak boson fusion.  In this case, though one has only weak-scale production cross sections, there is a large branching fraction into photons, therefore one may expect the spin two state to produce a significant number of diphoton events with a total cross section similar to that of the Higgs.

As the coupling to SU(3) is increased, gluon fusion becomes the dominant production mechanism of the spin two state and the production cross section increases dramatically .  However, once the tensor can decay into pairs of massless gluons they quickly begin to eat up the particle's branching fraction. In the appendix we have included some of the details of the calculation of the spin two state's branching fraction.   Figure 6b gives a contour plot of the spin two state's branching fraction into photon pairs in $k_1^{*}$ ,$k_2^{*}$ space - the space of effective couplings to U(1) and SU(2) gauge groups for fixed value of SU(3)coupling.

Such a state may fake a Higgs-like diphton signal.  Figure 6a shows the region of parameter space in which a 125 GeV spin two object may give an signal in the di-photon channel which is between .8 and 2 times that of a SM Higgs.  The separate shaded regions show the allowed region for different values of effective coupling to SU(3).  One sees that there is allowed parameter space where this coupling is 0 and the production and decay of the spin two particle follows only from electroweak processes.  One also sees that as the effective coupling to SU(3) is increased, the overall production cross-section of the spin two state increases greatly.  In this case, in order to not overproduce di-photon events, the overall branching fraction into photons must be small, meaning that the allowed parameter space is in regions with small effective couplings to $SU(2)\times U(1)$ gauge groups.

\section{Conclusions}

We have shown that a variety of effective operators may produce Higgs-like di-boson signals at LHC.  Scalar, pseudo-scalar and tensor particles which are uncharged under SM gauge groups  may couple to SM gauge bosons.  The new effective operators allow for states to be produced either through gluon fusion or processes such as weak boson fusion.  One may use current Higgs searches to place bounds on the parameter space of these models, in addition, these new states may have a large enough production cross section and branching fraction into di-photons to account for the 125 GeV excess in the di-photon channel at LHC.

Further areas of study may involve building completions of these models, such a task may or may not be difficult to achieve depending on the scenario.  In the case of dimension 5 operators which couple a scalar or pseudoscalar state to the SM gauge bosons, completions may simple.  The scalar may have couplings to TeV scale fermions charged under the SM gauge groups.  Spin two state may be new light composite states.  Any completion would likely involve new particles between 1 and 10 TeV which interact with the SM gauge groups and should have experimental consequences for searches conducted at 14 TeV.

\subsection{Acknowledgements}
Thanks to Arvind Rajaraman, Yuri Shirman and Daniel Whiteson for useful discussions and motivation.


\subsection{Branching Fraction Calculations}
\subsubsection{Standard Model Singlet}
The 3-point interaction from \eqref{eq:SMSinglet} that contributes to the decay of the scalar has the form
\bea
\mathcal{L} = 2\frac{k_i}{\Lambda}S\left(\partial_{\mu}A_{\nu}\partial^{\mu}A^{\nu}-\partial_{\mu}A_{\nu}\partial^{\nu}A^{\mu}\right)
\eea
This vertex has the Feynman rule $-ig_{SA_1A_2}(p_1\cdot p_2 g_{\mu\nu} - p_{1\mu}p_{2\nu})$, where $p_1 \text{ and }p_2$ are the momentum of the gauge bosons.

The decay width into massless states is straightforward to calculate.  We find
\bea
\Gamma(S \rightarrow \gamma\gamma) &=& \frac{g_{S\gamma\gamma}^2}{4\pi}m_s^3 \\
\Gamma(S \rightarrow gg) &=& \frac{2g_{Sgg}^2}{\pi}m_s^3 \\
\Gamma(S\rightarrow Z\gamma) &=& \frac{2g_{SZ\gamma}^2}{\pi}m_s^2\sqrt{m_s^2-m_Z^2}
\eea
The decay into Ws and Zs is a little more involved.  At least one of the massive gauge bosons will be off shell and decay
into fermion pairs ala \cite{Rizzo:1980gz}.  The matrix elements for these two cases are
\bea
|M|^2_{ZZ^*} = 24\pi\frac{\Gamma_Z}{m_Z}\frac{g_{SZZ}^2}{(q^2-m_Z^2)^2+\Gamma_Z^2m_Z^2}\frac{(q\cdot p_3)^2}{m_Z^2}\left[2(p_1\cdot p_3)(p_2\cdot p_3)+m_Z^2(p_1\cdot p_2)\right] \\
|M|^2_{WW^*} = 48\pi\frac{\Gamma_W}{m_W}\frac{g_{SWW}^2}{(q^2-m_W^2)^2+\Gamma_W^2m_W^2}\frac{(q\cdot p_3)^2}{m_W^2}\left[2(p_1\cdot p_3)(p_2\cdot p_3)+m_W^2(p_1\cdot p_2)\right]
\eea
where $p_1 \text{ and } p_2$ are the momentum of the fermions, $p_3$ the momentum of the final state gauge boson, and $q=p_1+p_2$ is the momentum of the off shell gauge boson.  In terms of invariant mass, the decay width to massive gauge boson pairs becomes
\bea
\Gamma(S \rightarrow Z f \bar{f})&=&\frac{3g_{SZZ}^2}{256\pi^2}\frac{\Gamma_Z}{m_s^3m_Z}\int dm_{12}^2 dm_{23}^2|M|^2 \\
\Gamma(S \rightarrow W f \bar{f'})&=&\frac{3g_{SWW}^2}{128\pi^2}\frac{\Gamma_W}{m_s^3m_W}\int dm_{12}^2 dm_{23}^2|M|^2\\
|M|^2&=& \frac{1}{(q^2-m_A^2)^2+\Gamma_A^2m_A^2}\frac{(m_S^2-m_A^2-m_{12}^2)^2}{m_A^2}\left[(m_S^2-m_{12}^2-m_{23}^2)(m_{23}^2-m_A^2)+m_A^2m_{12}^2\right]
\eea

\subsubsection{Pseudoscalar}
The pseudoscalar fake Higgs can decay into pairs of gluons, Z's, W's, photons and to Z plus photon.  The piece of the Lagrangian which contributes to these decays has the form
\bea
S\epsilon^{\mu\nu\rho\sigma}\partial_\mu A_{1\nu} \partial_\rho A_{2\sigma}
\eea
The Feynman rules for these vertices are given by $-ig_{SA_1A_2}\epsilon^{\mu\nu\rho\sigma}p_{1\mu}p_{2\rho}$ where the prefactors are again given in Section 3.  The decay width to massless pairs and to Z plus photon are straightforward to calculate.  They are
\bea
\Gamma(S \rightarrow \gamma\gamma) &=& \frac{g_{S\gamma\gamma}^2}{4\pi}m_s^3 \\
\Gamma(S \rightarrow gg) &=& \frac{2g_{Sgg}^2}{\pi}m_s^3 \\
\Gamma(S\rightarrow Z\gamma) &=& \frac{2g_{SZ\gamma}^2}{\pi}m_s^2\sqrt{m_s^2-m_Z^2}
\eea
For the WW and ZZ channel, at least one of the massive gauge bosons will decay to a pair of fermions.  In terms of momentum, the square of the matrix elements are
\bea
|M|^2_{WW^*}=24\pi\frac{\Gamma_W}{m_W}\frac{g_{SWW}^2}{(q^2-m_W^2)^2+\Gamma_W^2m_W^2}\left[2q^2(q\cdot p_3)^2 - m_W^2q^4 - 4q^4(p_1\cdot p_3)(p_2 \cdot p_3)\right] \\
 |M|^2_{ZZ^*} =12\pi\frac{\Gamma_Z}{m_Z}\frac{g_{SZZ}^2}{(q^2-m_Z^2)^2+\Gamma_Z^2m_Z^2}\left[2q^2(q\cdot p_3)^2 - m_Z^2q^4 - 4q^4(p_1\cdot p_3)(p_2 \cdot p_3)\right]
 \eea
We do the 3 body phase space integral in terms of invariant masses, $m_{12} \text{ and } m_{23}$.  For both cases, the square of the matrix element is just a polynomial in $m_{23}^2$ and so this integral can be done analytically.  Because of the limits of integration on the $m_{23}$ integral, we do the $m_{12}$ integral numerically.  In the rest frame of the decaying particle, the decay widths into $Wff'$ and $Zff$ are
\bea
\Gamma(S \rightarrow W f \bar{f'})&=&\frac{3g_{SWW}^2}{32\pi^2}\frac{\Gamma_W}{m_Wm_S^3}\int dm_{12}^2 dm_{23}^2 M^2\\
\Gamma(S \rightarrow Z f f)&=&\frac{3g_{SZZ}^2}{64\pi^2}\frac{\Gamma_Z}{m_Zm_S^3}\int dm_{12}^2 dm_{23}^2 M^2 \\ \\
M^2 &=& \frac{m_{12}^2}{(m_{12}^2-m_A^2)^2+\Gamma_A^2m_A^2}\left[\frac{1}{2}(m_s^2-m_A^2-m_{12}^2)^2-(m_s^2-m_{12}^2-m_{23}^2)(m_{23}^2-m_A^2)-m_A^2m_{12}^2\right]
\eea
and $m_A$ is either the W or Z mass.
\subsubsection{Tensor}

The completeness condition for the polarization tensors for $T_{\mu\nu}$ is
\bea
\sum_s \epsilon_{\mu\nu}^s \epsilon_{\rho\sigma}^{s*} = B_{\mu\nu,\rho\sigma}(p)
\eea
The spin two pseudo-higgs can decay into pairs of gluons, photons, $Z\gamma$, $WW*$, and $ZZ*$.  The massless case
and the $Z\gamma$ case can be calculated analytically.  We find
\bea
\Gamma(S \rightarrow \gamma\gamma) &=& \frac{g_{S\gamma\gamma}^2}{4\pi}m_s^3 \\
\Gamma(S \rightarrow gg) &=& \frac{2g_{Sgg}^2}{\pi}m_s^3 \\
\Gamma(S\rightarrow Z\gamma) &=& \frac{g_{SZ\gamma}^2}{16\pi m_s^2}\sqrt{m_s^2+m_Z^2}M_{Z\gamma}^2 \\
M_{Z\gamma}^2 &=& \frac{1}{3}\left(4\frac{m_Z^8}{m_s^4}-26\frac{m_Z^6}{m_s^2}+52m_Z^4\right)-
\left(\frac{4}{3}\frac{m_Z^6}{m_s^4}-8\frac{m_Z^4}{m_s^2}-\frac{4}{3}m_Z^2\right)(k_1\cdot k_2) \\
&-&\left(\frac{32}{3}\frac{m_Z^4}{m_s^4}-\frac{22}{3}\frac{m_Z^2}{m_s^2}-\frac{8}{3}\right)(k_1\cdot k_2)^2 -
\left(\frac{56}{3}\frac{m_Z^2}{m_s^4}-\frac{4}{3}\frac{1}{m_s^2}\right)(k_1\cdot k_2)^3-
16\frac{1}{m_s^4}(k_1\cdot k_2)^4 \\
k_1\cdot k_2 &=& m_s\sqrt{m_Z^2+m_s^2}+m_s^2
\eea
One of the massive gauge bosons will decay into fermion pairs for the massive diboson case.  The matrix elements squared
in this case are
\bea
|M|_{WW*}^2 &=& 6\pi\frac{\Gamma_W}{m_W}\frac{g_{SWW}^2}{(q^2-m_W^2)^2+\Gamma_W^2m_W^2}\left(k_1^ak_2^\alpha-k_1\cdot k_2 g^{\alpha a}+k_1^\alpha k_2^a\right)\left(-g^{\sigma s}+\frac{k_3^\sigma k_3^s}{m_W^2}\right) \\
\left(q\cdot k_3C_{\mu\nu,\alpha\sigma}+D_{\mu\nu,\alpha\sigma}\right)\left(q\cdot k_3 C_{mn,as}+D_{mn,as}\right) \\
|M|_{ZZ*}^2 &=& 3\pi\frac{\Gamma_Z}{m_Z}\frac{g_{SZZ}^2}{(q^2-m_Z^2)^2+\Gamma_Z^2m_Z^2}\left(k_1^ak_2^\alpha-k_1\cdot k_2 g^{\alpha a}+k_1^\alpha k_2^a\right)\left(-g^{\sigma s}+\frac{k_3^\sigma k_3^s}{m_Z^2}\right) \\
\left(q\cdot k_3C_{\mu\nu,\alpha\sigma}+D_{\mu\nu,\alpha\sigma}\right)\left(q\cdot k_3 C_{mn,as}+D_{mn,as}\right)
\eea
where $k_1 \text{ and } k_2$ are the four momenta of the final state fermions, $k_3$ is the four momentum of
the final state gauge boson and $q$ is gauge boson propagator momentum.  The index contraction for these matrix elements was done using FORM \cite{J.A.M.Vermaseren:2000}.  The three body phase space integral, done in terms of invariant
mass, was computed numerically.

\end{document}